# Earth's polar night boundary layer as an analogue for dark side inversions on synchronously rotating terrestrial exoplanets

Short title: Boundary layers on synchronously rotating planets

M. M. Joshi[1,2], A. D. Elvidge[1], R. Wordsworth[3,4], D. Sergeev[5]

[1] School of Environmental Sciences, University of East Anglia, UK
[2] Climatic Research Unit, School of Environmental Sciences, University of East Anglia, UK
[3] School of Engineering and Applied Sciences, Harvard University, USA
[4] Department of Earth and Planetary Sciences, Harvard University, USA
[5] College of Engineering, Mathematics and Physical Sciences, University of Exeter, UK

Postal Addresses:
Manoj Joshi, Andrew Elvidge: School of Environmental Sciences, University of East Anglia, Norwich Research Park, Norwich, NR4 7SQ, United Kingdom
Robin Wordsworth: 26 Oxford Street, School of Engineering and Applied Sciences, Harvard University, Cambridge MA 02138, USA
Denis Sergeev, CEMPS, Laver Building, Streatham Campus, University of Exeter
North Park Road, Exeter, EX4 4QF, UK

Correspondence email: m.joshi@uea.ac.uk

## ABSTRACT

A key factor in determining the potential habitability of synchronously rotating planets is the strength of the atmospheric boundary layer inversion between the dark side surface and the free atmosphere. Here we analyse data obtained from polar night measurements at the South Pole and Alert Canada, which are the closest analogues on Earth to conditions on the dark sides of synchronously rotating exoplanets without and with a maritime influence, respectively. On Earth, such inversions rarely exceed 30 K in strength, because of the effect of turbulent mixing induced by phenomena such as so-called mesoscale slope winds, which have horizontal scales of 10s to 100s of km, suggesting a similar constraint to near-surface dark side inversions. We discuss the sensitivity of inversion strength to factors such as orography and the global-scale circulation, and compare them to a simulation of the planet Proxima Centauri b. Our results demonstrate the importance of comparisons with Earth data in exoplanet research, and highlight the need for further studies of the exoplanet atmospheric collapse problem using mesoscale and eddy-resolving models.

## 1. INTRODUCTION

M-dwarfs form up to 80% of main-sequence stars, and have the most observable exoplanets because of factors such as these stars' relatively small sizes, the relatively short periods of planets in their habitable zones, and the relatively high transit probability of such planets (e.g. Seager 2010, Anglada-Escudé et al. 2016, Gillon et al. 2016). As a result, questions related to their habitability have been examined for decades (Kasting et al. 1993, Tarter et al. 2007, Kite et al. 2011, Shields et al. 2016, Turbet et al. 2016). A key property of many planets in the habitable zone of M-stars is that they are quite likely to be synchronously rotating with respect to their parent stars, i.e. with one side always facing the star while the other faces away (e.g. Dole, 1964); the likelihood of such a state existing is dependent on factors such as stellar mass, planetary semi-major axis and atmospheric mass (Leconte et al 2015). The phenomenon of synchronous rotation implies far greater spatial gradients in time-averaged stellar heating than are experienced on Earth (Kasting et al. 1993), implying significant day-night temperature gradients. Several studies have shown the critical role of atmospheric heat transport in determining the strength of this gradient (Joshi et al. 1997, Joshi 2003, Merlis and Schneider 2010, Leconte et al. 2013, Wordsworth 2015, Koll & Abbot 2016). More recent studies have shown that the effect of ocean heat transport might also be significant (Hu and Yang 2014).

The transport of heat from day to night is not the only factor determining the surface temperature of the dark side because in these locations, the surface is likely to be much colder than the overlying atmosphere, especially if the surface is land (Joshi et al. 1997, Wordsworth 2015). When the underlying surface is colder than the atmosphere, near-surface temperature gradients can be much larger than when the surface is warmer than the atmosphere. Such so-called inversions occur because the turbulent convective motions that mix heat in the vertical are almost absent when the vertical atmosphere structure has a positive gradient of potential temperature with height, usually denoted as a "stable atmosphere" (e.g. Salmond and McHenry 2005). The strength of such inversions can be quantified from planetary boundary layer theory (e.g. Garratt 1994), but effects such as topographically driven circulations and associated turbulence, or the strength of the background flow, can all complicate the application of this theory to real planetary scenarios (e.g. Physik 1976, Haberle et al. 1993, Orr et al. 2014).

The first studies of synchronous rotators had relatively small surface inversions (e.g. Joshi et al. 1997), but this was likely because of two factors: firstly, the vertical and horizontal resolution used in global modelling studies mean that small-scale momentum and tracer exchange must be parameterized rather than explicitly resolved, and such parameterizations vary widely in complexity (e.g. James and Gray 1986, Forster et al 2000, Mellor and Yamada 1982); secondly, the low vertical resolution of early studies (e.g. Joshi et al. 1997) meant that near-surface inversions would not be well-resolved even when predicted by the model's parameterization. More recent studies carried out using higher vertical resolution and more sophisticated planetary boundary layer schemes have suggested that surface inversions might be much larger in magnitude (Wordsworth 2015), implying that condensation of gases such as $CO_2$ can occur even with a relatively warm free atmosphere above. However, the situation is complicated by the fact that while global-scale climate models can generally parameterize the effects of turbulence on scales of less than 1 km on surface-atmosphere exchange, they do not parameterize mesoscale topographically-driven flows having scales of 10-1000 km (Joshi et al. 1997, Wordsworth et al. 2015), and the resulting turbulent eddies, which mix heat in the vertical over land. Such an omission suggests that even GCM studies with state-of-the-art boundary layer representations and high resolution might overestimate inversion strength when run with "flat" or "oceanic" boundary conditions.

The effects of topographical circulations on near-surface dark side inversions are important for two reasons. Firstly, if one assumes that on a given exoplanet, landlocked regions on the dark side are, all other things being equal, likely to be colder than nearby oceanic regions which have heat transported to them from the dayside oceans (Hu and Yang 2014), then landlocked regions are likeliest to first undergo surface $CO_2$ condensation if a synchronously rotating planet cools for any given reason. Secondly, even relatively flat regions of the Earth do have mesoscale heterogeneity in topography which can induce weak circulations (Elvidge et al. 2019). It is therefore likely that the coldest regions of terrestrial exoplanets will have some sort of topographically-related circulation causing turbulent mixing, and mitigating the strength of dark side near-surface inversions.

While resolving the issue of dark side inversions is clearly not possible with observations of exoplanets, three potential avenues of exploration do exist. Observations of analogue situations on Earth, simple theoretical modelling, and more complex numerical modelling of horizontal scales of up to 1000 km can all be utilised to understand the issue. Regarding the first avenue, the Earth's South Pole is the coldest region on Earth, and experiences well over 90 days of total darkness each year. During this time, the atmosphere and surface can cool effectively to space, resulting in an approximate equilibrium between heat transport from lower latitudes and radiative cooling to space, since the radiative relaxation timescale of the Earth's atmosphere is far less than the approximately 90-day duration of polar night (Tompkins and Craig 1998). Such a situation is the closest approximation on Earth to the energetic balance predicted on the dark side of a synchronously rotating planet. Alert Canada has a similar radiative balance, but, due to a stronger maritime influence, is warmer. As well as providing a useful maritime counterpoint, observations here can help to test the robustness of the inversion strengths observed at the South Pole. In this paper we examine the strengths of wintertime inversions in observations of the South Pole and Alert Canada, and in simplified models of flows over topographical slopes. We compare these results with a climate model simulation of Proxima Centauri b, and discuss the impacts of our results on constraining the magnitude of inversions in the boundary layers of synchronously rotating planets, and what this implies for potential atmospheric collapse and habitability.

## 2. METHOD

The data series spans 10 years (2008-2017) of winter radiosonde data from the South Pole research station during the June-July-August southern winter season (JJA) and from Environment and Climate Change Canada's weather station at Alert, Nunavut, Canada during the December-January-February northern winter season (DJF). The South Pole experiences the longest wintertime nights on Earth, whilst Alert, lying at 82.5N, is one of the northernmost points of land on Earth, ~830 km south of the North Pole, and so also experiences a long polar night. The South Pole station lies on the Antarctic Plateau at an altitude of 2839 m above sea level, whilst Alert station is close to the coastline, lying 76 m above sea level. Although the Arctic radiosonde data are much lower in vertical resolution, there is usually an initial reading at ~10 m, plus the surface reading. Launches at Alert are at 00 and 12 UTC, and at S Pole at 00 UTC and, intermittently, at 12 UTC.

In regions of sloping terrain capped by a surface inversion, cold and dense flows known as katabatic winds are generated, driven downslope by a combination of gravity and horizontal pressure gradient force and confined to near-surface by the inversion. The simplest analytical model for these slope winds is that of Prandtl (1952), which represents the advection and turbulent diffusion of momentum and sensible heat within the katabatic flow over uniformly sloping terrain by assuming a one-dimensional steady-state balance between buoyancy and turbulent friction. In this study we have used the Prandtl model in its classic, linear form using a constant eddy heat conductivity to derive wind profiles of idealised, steady state katabatic flows for varying terrain slope $\alpha$ and surface inversion strength as follows:

$$U(z) = -C\mu \exp\left(\frac{-z}{h_p}\right)\sin\left(\frac{z}{h_p}\right),$$

where $C$ is the potential temperature perturbation at the surface (negative under katabatic conditions), $\mu = \sqrt{g/(\Theta_0 \Gamma_{bkg} Pr)}$, $h_p = \sqrt{2}/\sigma$ is the characteristic depth of the Prandtl layer and $\sigma^2 = \Gamma_{bkg}\sin(\alpha)\,g/(\Theta_0 K\sqrt{Pr})$. All constant parameters are set to typical Prandtl model values following Grisogono et al. (2015), as follows: eddy heat conductivity $K = 0.06$, Prandtl number $Pr = 2$, reference temperature $\Theta_0 = 273.15$ K, and background or free-flow potential temperature gradient $\Gamma_{bkg} = 3$ K km$^{-1}$ (this value is supported by our South Pole observations above the inversion layer). To relate the Prandtl model to our observed and model profiles, we define $C$ as follows:

$$C = d_{inv}(\Gamma_{bkg} - \Gamma_{inv}),$$

where $\Gamma_{inv}$ is the strength of the surface inversion in terms of the vertical gradient in potential temperature and $d_{inv}$ is the depth of the surface inversion, which we have set to 100 m; the typical depth of katabatic flows (Renfrew and Anderson, 2002).

The climate model used in this paper is the Global Atmosphere 7 (GA7.0) configuration of the Met Office Unified Model (UM) (Walters et al. 2019). Parameterized processes include subgrid-scale turbulence, convection, $H_2O$ cloud and precipitation formation (with prognostic ice and liquid phases), and radiative transfer (the SOCRATES correlated-k scheme). Full details of UM dynamics and physics can be found in Walters et al. (2019) and references therein. The orbital characteristics of the nearest confirmed terrestrial planet, Proxima Centauri b (Anglada-Escudé et al. 2016) are used, with the stellar spectrum of its host star taken from the BT-Settl model (Rajpurohit et al. 2013). The planet is assumed to be in 1:1 spin:orbit resonance. All simulations have a $N_2$-dominated atmosphere with trace amounts of $CO_2$ and $H_2O$ amounting to a mean surface pressure of 1 bar.

The UM is used in so-called 'aquaplanet' mode, assuming a so-called 'swamp' ocean having zero horizontal heat flux, and a depth of 2.4 m. Such a configuration, which has been used in previous studies of this type (e.g. Boutle et al. 2017), makes no assumptions about the nature of the ocean circulation on an exoplanet, which can be sensitive to factors such as continental configuration (e.g. Cullum et al. 2016). Simulations are performed with a horizontal grid spacing of 2.5° in longitude and 2.0° in latitude, with 38 vertical levels quadratically stretched from the surface to the model top, located at approximately 40 km height. The results presented are averages over the last 1000 days of a simulation that is 2000 Earth days long, and the runs referred to as "UMPb". The criterion for the UM reaching equilibrium is when the top-of-atmosphere radiation fluxes, and the global hydrological cycle (precipitation minus evaporation at the surface), are in balance.

3. RESULTS

Figure 1 shows the state of the atmospheric boundary layer during polar night at the South Pole and Alert over a decade of observations and on the dark side of Proxima b over the 1000-year UMPb simulation. Figure 1 (a), which shows how potential temperature $\theta$ varies in height from the surface, shows that deep inversions are typical; note the large increase in mean potential temperature of 20 K from the surface to 300 m height at the South Pole (blue curve). The average inversion strength at Alert Canada (red curve) is much smaller – approximately 7 K in terms of potential temperature – reflective of Alert's maritime influence. Average potential temperature inversions at the coldest dark-side location in run UMPb (black curve) are much stronger than the polar observations, being typically 40 K within the lowest 100 m of the surface, rising to 50 K within 300 m of the surface. Such strong inversions are similar in magnitude to those described by Wordsworth et al (2015).

Another method of diagnosing the deep inversions in observations and model is to calculate static stability $N^2$. Figure 1 (b) shows that static stability at the South Pole (blue curve) peaks at $5\times10^{-3}$ s$^{-2}$ within 50 m of the surface, reducing to $1.5\times10^{-3}$ s$^{-2}$ by 300 m altitude. By contrast $N^2$ at Alert Canada is mostly lower, being $<1\times10^{-3}$ s$^{-2}$ (red curve) except near the surface, with the mean stability below a height of 20 m being slightly greater at Alert. Again, run UMPb (black curve) displays much higher static stabilities than either polar location, reaching values of $5\times10^{-2}$ s$^{-2}$ at the lowest model level.

Figure 1 (c) shows windspeed ($v$), which at the South Pole (blue curve) has a mean magnitude of ~6 ms$^{-1}$ at the surface, rising to 10 ms$^{-1}$ above the boundary layer. $v$ does vary strongly in time, and is closely related to $N^2$, since weak winds inhibit vertical mixing, allowing large inversions to build. However, over sloped ground, increasing static stability induces near-surface winds via the katabatic effect. Generally, $v$ at Alert is weaker than at the South Pole, having a magnitude of approximately 2-3 ms$^{-1}$ at near-surface, rising to 5-6 ms$^{-1}$ above the boundary layer. This difference is a possible cause for the stronger mean stability observed below 20 m at Alert compared to the South Pole. Run UMPb (black curve) displays winds that are typical 5 ms$^{-1}$, reducing to approximately 1 ms$^{-1}$ in the lowest 50 m of the atmosphere. Such weak winds are consistent with the very large mean inversion shown in Figure 1 (b) (black curve).

In order to understand the potential implications of the above observations to the dark side of synchronously rotating exoplanets, near-surface $N^2$ is compared with $v$, because the background flow on an exoplanet is thought to vary strongly with planetary rotation rate (Haqq-Misra et al. 2018), influencing the total windshear and near-surface mixing on the dark side of an exoplanet (see below). Figure 2 (a) shows $N^2$ between the surface and 10 m height at the South Pole vs $v$. Generally, as $v$ decreases, both the average and the spread in $N^2$ increase. When $v$ is very small, the average $N^2$ in the lowest 10 m of the atmosphere is typically of order 0.01 s$^{-2}$, with a median value of 0.08 s$^{-2}$ for wind speeds below 3 m s$^{-1}$, which is equivalent to an inversion of approximately 2 K in the lowest 10 m of the atmosphere. A similar relationship can be found when looking at a deeper section of the atmosphere. Figure 2 (b) shows that average values of $N^2$ in the weakest winds differ little from those in the lowest 10 m, with a median value again of 0.08 s$^{-2}$ in the lowest 100 m when $v$ falls below 3 ms$^{-1}$. Such inversions are approximately equivalent to 20 K in potential temperature $\theta$. The strongest surface inversion in the lowest 100 m appearing in our observations is 30 K in terms of potential temperature difference with respect to the surface, occurring at the South Pole. The 10-m wind speed during this case was 2 ms$^{-1}$.

A key question that follows on from the above results is the applicability or generality of the results obtained above to synchronously rotating planets in general. The South Pole itself typically experiences winds of 1-12 ms$^{-1}$ (95 % of the time; see Figure 2), though the coastal margins of Antarctica typically experience much stronger winds because of katabatic slope flows. The mean planetary-scale background flow on an exoplanet has the potential to be much weaker than 1 ms$^{-1}$, depending on the planet's rotation rate, potentially allowing the existence of very strong inversions on its dark side. Conversely, even relatively flat regions on Earth experience slope winds, driven by density gradients between highland and lowland regions, which cause near-surface turbulence and erode large inversions.

Such a scenario is explored using the Prandtl model described in section 2. Figure 3 shows wind speeds for a number of slopes and near-surface inversion strengths. The steepest slope is shown in yellow and the shallowest shown in purple. It should be noted that in reality, there is a negative feedback between $v$ and inversion strength, since the former generates turbulence that lowers the latter, which in turn lowers the density gradients that drive slope winds and the magnitude of *v*. Nevertheless, the Prandtl model can give an indication of the maximum strength of winds generated by inversions over sloping terrain. Figure 3 shows that *v* typically increases as slope gradient increases, consistent with the idea of larger slopes leading to larger buoyancy forcing of winds. Even for shallow slopes of 0.2 m km$^{-1}$, representative of the flattest terrain on Earth (e.g. the Antarctic plateau) for horizontal distances of O (10 km) (Elvidge et al. 2019), an inversion of strength equal to that observed to be typical at the South Pole (blue vertical line) confined to the lowest 100 m of the atmosphere will cause a slope wind of strength 1 ms$^{-1}$.

## 4) DISCUSSION

While the results presented in Figure 3 are idealised, they do suggest that even small slopes of magnitude 0.2 m km$^{-1}$ can generate buoyancy-driven currents of ~1 ms$^{-1}$ in strength under static stabilities typical of Earth's polar regions. An examination of Figure 2 suggests that even with such weak background winds, sufficient turbulence is created near the surface to prevent inversions that exceed the typical values observed at the South Pole. To put a slope magnitude of 0.2 m km$^{-1}$ or 0.011° into context, such a value is approximately equal to a change in altitude of 200 m or approximately 600 feet over a distance of 1000 km or 600 miles, which is flatter than the US Great Plains region, or indeed most regions of Mars, including its northern plains (Aharonson et al. 2001).

A simplified framework for limiting the strengths of near-surface inversions does allow for a parameterization of such phenomena. In general circulation models (GCMs), limiting inversion strength may be achieved by allowing a minimum amount of surface-atmosphere mixing under stable conditions. Alternatively, the maximum inversion strength in the lowest 100 m occurring in our observations of approximately 30 K could be used to determine dark side surface temperatures when atmospheric temperatures are calculated in simpler energy balance models (EBMs) that parameterize rather than simulate planetary-scale circulations (e.g. Kite et al. 2011). In such cases though, the generality of the above approach may be limited if the planetary atmosphere in question is very different to that of Earth because, for instance, the surface pressure is far lower than 1000 hPa.

In thinner atmospheres than Earth's, GCM simulations show that inversions increase in strength significantly as pressure decreases, presumably because an optically thinner atmosphere allows greater surface cooling to space (Wordsworth et al. 2015, Figure 2); such an effect implies a positive feedback between inversion strength and surface pressure. However, observations and models of the Martian atmosphere suggest that while strong night-time near-surface inversions occur as part of the diurnal cycle (Haberle et al 1993) - a result of the very low thermal inertia of the surface - they are actually quite rare over most of the polar cap region during polar night, possibly because they are mixed away by mesoscale circulations driven by slope winds (Spiga et al. 2011), or thermal contrasts between the ice cap itself and bare soil at lower latitudes (Siili et al. 1999). Such phenomena imply the potential for a negative feedback that hinders atmospheric collapse on a synchronously rotating planet: as a $CO_2$ ice cap builds up on the cold dark side, its presence induces slope winds that reduce inversion strength, warming the surface, and sublimating away the ice.

Another potential negative feedback hindering atmospheric collapse might be the build-up of atmospheric $CO_2$ (and hence atmospheric mass) associated with lower weathering rates on a cold planet. Such a process would only significantly affect the dark side if atmospheric $CO_2$ accumulated to partial pressures that were a significant fraction of the whole atmosphere. The strength of such feedbacks would therefore be dependent on atmospheric composition, as well as weathering rates on the starlit side of the planet (e.g. Edson et al 2012). It should be noted that such a feedback would be unlikely on a planet with a collapsed atmosphere, as $CO_2$ produced by volcanism would be deposited as ice on the dark side, rather than building up in the atmosphere.

Most of the results in this paper have concentrated on the South Pole because the Alert base region is a maritime area. An analogous region on a synchronously rotating exoplanet might be an area where oceanic heat fluxes from dayside to nightside play a significant role in controlling near surface temperature, so surface temperatures in such an area might be expected to be much higher than in the interior of a continent on the dark side of an exoplanet. Such a situation might be examined in future using GCMs with a dynamic ocean component. In addition, turbulent fluxes associated with mesoscale land-sea breezes (Physik 1976) would also reduce inversion strength over cold land-covered regions. Conducting idealised studies of dark side near-surface conditions using regional circulation models which represent boundary layer processes better than GCMs, and explicitly represent mesoscale circulations such as slope flows (e.g. Orr et al., 2014) or land-ocean breezes, under conditions of varying surface pressure and rotation rates, would help to reduce such uncertainties.

## 4) CONCLUSIONS

Measurements of the polar night boundary layer on Earth, and simplified modelling of slope-driven flows, suggest that inversions in temperature rarely exceed 30 K in the lowest 100m of the atmosphere. Such inversions are lower than might be expected from global circulation models with flat bottom boundary conditions and parameterized boundary layer mixing, suggesting that mesoscale circulations of horizontal scale 100-1000 km, most likely driven by topographical variations, are causing sufficient mixing to limit inversion strength. In the scenario of the dark side of a synchronously rotating planet, such mixing would inhibit the surface from cooling compared to the relatively warm atmosphere aloft.

Most of the results in this paper have concentrated on the South Pole because the Alert base region is a maritime area. An analogous region on a synchronously rotating exoplanet might be an area where oceanic heat fluxes from dayside to nightside play a significant role in controlling near surface temperature, so surface temperatures in such an area might be expected to be much higher than in the interior of a continent on the dark side of an exoplanet. In addition, turbulent fluxes associated with mesoscale land-sea breezes (Physik 1976) would also reduce inversion strength over cold land-covered regions. Conducting idealised studies of dark side near-surface conditions using regional circulation models which represent boundary layer processes better than GCMs, and explicitly represent mesoscale circulations such as slope flows (e.g. Orr et al., 2014) or land-ocean breezes, under conditions of varying surface pressure and rotation rates, would help to reduce such uncertainties.

In this paper we have attempted to place observationally-based limits on the strengths of near-surface inversions on the dark sides of synchronously rotating planets, as such phenomena will play a very important role in controlling the coldest surface temperatures, atmospheric collapse, and hence habitability, of such planets. While observations and simplified modelling can only be a guide to conditions on even somewhat Earth-like synchronously rotating worlds, they do show how well-chosen observations of the Earth system can shed light on understanding the potential habitability of terrestrial exoplanets.

**ACKNOWLEDGEMENTS**

We acknowledge the assistance of James Murphy. The authors appreciate the support of the University of Wisconsin-Madison Antarctic Meteorological Research Center in making available the South Pole radiosonde data, (NSF grant number ANT-1535632), and the Department of Atmospheric Science, University of Wyoming together with Environment and Climate Change Canada for the Alert, Canada radiosonde data. This work benefited from the 2018 Exoplanet Summer Program in the Other Worlds Laboratory (OWL) at the University of California, Santa Cruz, a program funded by the Heising-Simons Foundation. This work was partly supported by a Science and Technology Facilities Council Consolidated Grant (ST/R000395/1). We acknowledge use of the MONSooN system, a collaborative facility supplied under the Joint Weather and Climate Research Programme, a strategic partnership between the Met Office and the Natural Environment Research Council.

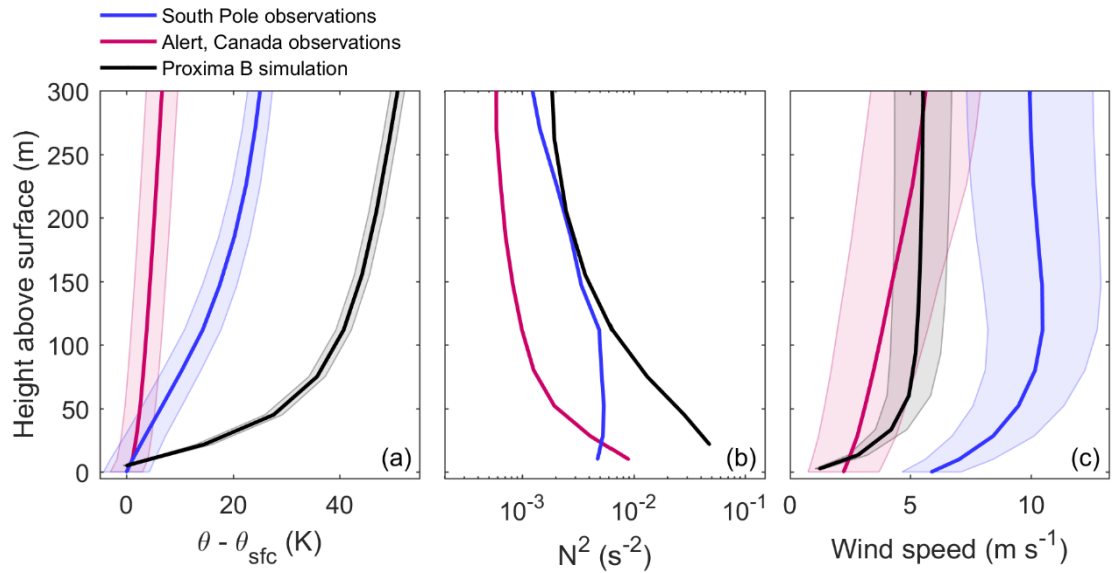

Figure 1: Meteorological variables during June-August in the lowest 300 m of the atmosphere observed at the South Pole from 2008 to 2017 (blue), observed at Alert Canada from December-February 2008-2018 (red) and simulated for the coldest region in the Proxima Centauri b simulation (UMPb) over 1000 days (black). (a) Mean (solid lines) and standard deviation (shading) of the potential temperature difference from the surface value $\theta$-$\theta_{sfc}$ (K); (b) mean of stability $N^2$ (s$^{-2}$); mean (solid lines) and standard deviation (shading) of windspeed $v$ (ms$^{-1}$). The mean surface temperatures at S Pole, Alert and UMPb are 240, 244 and 153 K, respectively.

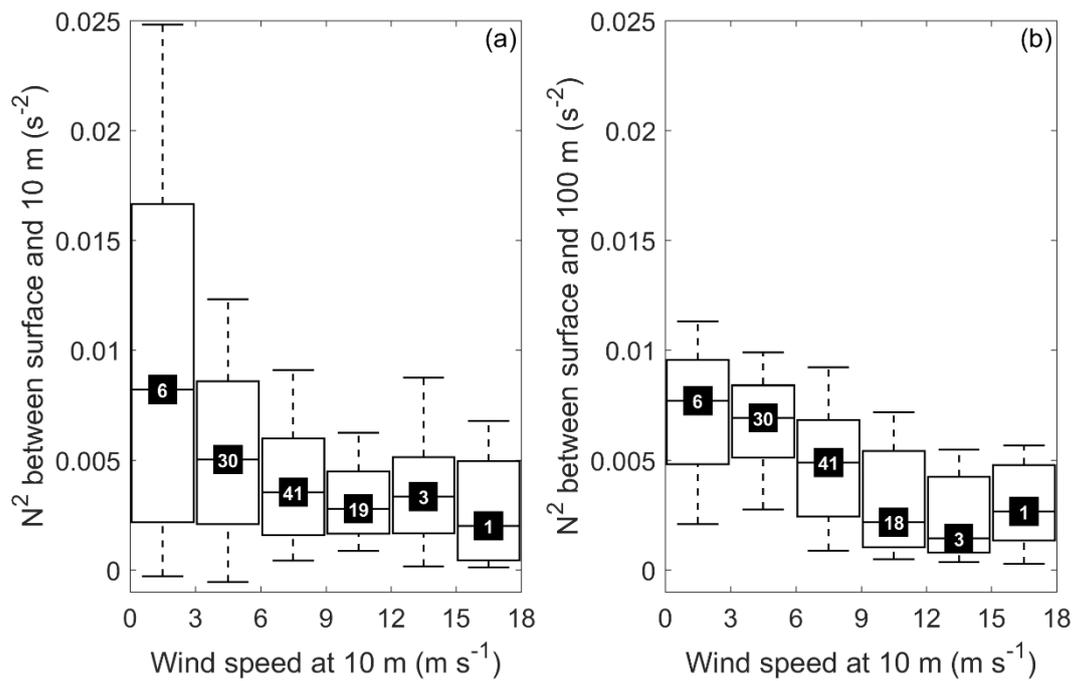

Figure 2. Box plots showing the median, interquartile range and 9th and 91st percentiles of (a) average $N^2$ ($s^{-2}$) between the surface and 10m altitude; and (b) average $N^2$ between the surface and 100m altitude, as a function of wind speed at 10 m altitude. The $N^2$ data have been grouped into wind speed bins of 3 m s$^{-1}$. The white numbers within black boxes centred on median values indicate the percentage of the total number of data points in each bin.

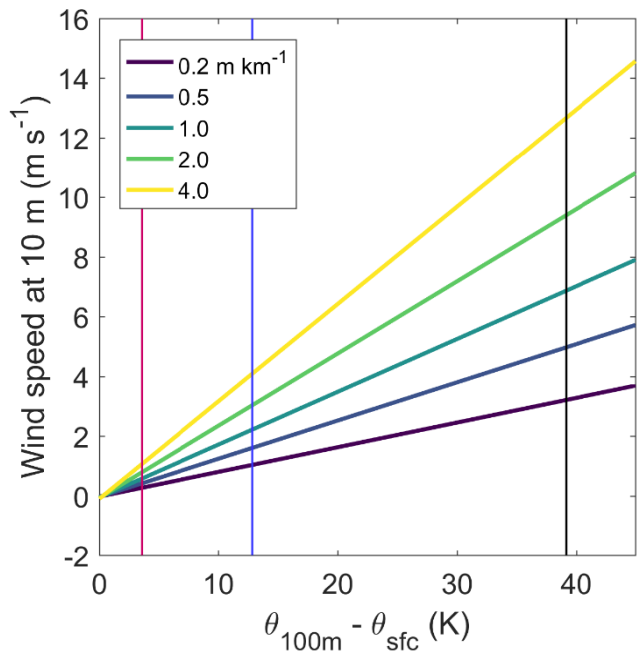

Figure 3. Prandtl model solution for 10m wind speed *v* (ms$^{-1}$) vs inversion strength in terms of the difference in potential temperature θ (K) between the surface and 100 m altitude for different idealised slopes. The blue, red and black vertical lines show the mean model θ difference in the South Pole observations, Alert observations and Proxima Centauri b simulation, respectively, between the surface and 100 m. Typical Antarctic plateau slopes vary from 0.2 to 4 m km$^{-1}$.